# Cloud Computing Research Challenges

Sultan Ullah
School of Computer and Communication Engineering
University of Science and Technology Beijing
Beijing, People Republic of China
sultan.ustb@yahoo.com

Zheng Xuefeng
School of Computer and Communication Engineering
University of Science and Technology Beijing
Beijing, People Republic of China
zxfxue@ies.ustb.edu.cn

*Abstract*— **In recent times cloud computing has appeared as a new model for hosting and conveying services over the Internet. This model is striking to business vendors as it eradicates the requirement for users to plan in advance, and it permits the organization to start from low level and then add more resources only if there is an increase in the service demand. Even though cloud computing presents greater opportunities not only to information technology industry, but every organization involved in utilizing the computing in one way or the other, it is still in infancy with many problems to be fixed. The paper discusses research challenges in cloud computing.**

## I. INTRODUCTION

Most of the computing and none computing organization needs to deal with large scale, vast amount of data, the demand of computing power far beyond the computing power of its own IT infrastructure, then you need to continue to increase system investment in hardware to achieve the scalability of the system due to the rapid growth of the internet era, information and data, science and engineering and commercial and industrial computing. In addition, due to the traditional parallel programming mode limitations of applications requires an objective and easy to learn, use and deployment of new parallel programming framework. In this case, in order to save costs and implement a system with large computing power the concept of cloud computing was introduced.

Cloud computing is one of the most important invention in the computing industry after the advent of personal computer and Internet. The transfer to cloud will be a major change in the computing industry. Cloud computing is a paradigm in which tasks are assigned to a combination of connections, software and services accessed over a network. it has raised an enormous interest both in academic and industry communities. This field is in infancy and going through an evolving phase that incorporates the evolutionary development of many existing computing technologies such as distributed services, applications, information and infrastructure consisting of pools of computers, networks, information and storage resources [1].

Cloud Computing as defined by the US National Institute of Standards and Technology (NIST) states that:"Cloud computing is a model for enabling convenient, on-demand network access to a shared pool of configurable computing resources (e.g., networks, servers, storage, applications, and services) that can be rapidly provisioned and released with minimal management effort or service provider interaction. This cloud model promotes availability and is composed of five essential characteristics, three delivery models, and four deployment models." [2]. Cloud computing has become a very attractive model because of its apparent economic and operational benefits.

The cloud provides super computer level power of computing to the end user. For example if we use a thin client or any other access point, user can reach to the cloud for resources whatever they need it. Therefore cloud computing can also be expressed as on demand computing. This enormous power is made possible through large – scale, distributed, cluster computing, often in association with virtualization software, for example Xen and parallel processing. To paraphrase Sun Microsystems' famous axiom, in cloud computing the network becomes the supercomputer [3].

The cloud presents the organizations with enormous opportunities and value, but despite of these huge benefits to the organizations, the main hurdle in adoption of cloud computing by most of the organizations is two main issues i.e Security and Privacy. It is shown by several surveys these issues will continue to keep some companies out of cloud computing [4]. Nevertheless, cloud computing appears to be an inevitable force because of its potential benefits.

When we shift our information into the cloud, we may no longer have control of it. The cloud provides us with the access to the data, but the most important thing is to ensure that only authorized entities have access to the data. It is important to understand how we can protect our data and resources from a security violation in the cloud that provides shared platforms and services [5].

In a cloud computing environment the power of computing is supplied by a large collection of data centers, which are typically installed with a huge number of servers [6] [7] . A typical four layer architecture is built by the authors. At the base layers there exist gigantic physical resources (storage servers and application servers) that power the data centers. These servers are transparently managed by the higher level virtualization services and toolkits that allow sharing of their capacity among virtual instances of servers [8].

## II. WHAT IS CLOUD COMPUTING

Cloud computing is an infrastructure that provides storage, processing and applications all as services. We could look at the cloud from two perspectives, one from user perspective and other from the service provider perspective. User perspective could be stated as to pay only for the services user has requested and utilized over the internet. User does not need to install any especial software to acquire all the services he wants; all he has to do is to have an internet connection and a web browser. Cloud service provider perspective could be stated as the cloud computing is a model that enables easy, on demand, automatic access to pooled resources which could be provisioned and de provisioned automatically without human intervention.

Cloud computing can be classified by the model of service it offers into one of three different groups. These will be described using the XaaS taxonomy, where "X" is Software, Platform, or Infrastructure, and the final "S" is for Service. It is important to note, that SaaS is built on PaaS, and the latter on IaaS. Hence, this is not an excluding approach to classification, but rather it concerns the level of the service provided.

Since cloud computing is an infrastructure, it is divided into three layers which are as under [1] [2] [3] [7]:

- Infrastructure as a Service (IaaS)
- Platform as a Service (PaaS)
- Software as a Service (SaaS)

### A. Infrastructure as a Service (IaaS)

The First layer is the IaaS. The capability provided to the customer of IaaS is raw storage space, computing, or network resources with which the customer can run and execute an operating system, applications, or any software that they choose. The cloud customer is not able to control the distribution of the software to a specific hardware platform or change parameters of the underlying infrastructure, but the customer can manage the software deployed.

Examples of IaaS providers include Amazon EC2 [13], GoGrid [14] and Flexiscale [15].

### B. Platform as a Service (PaaS)

Second layer is the PaaS. In the case of PaaS, the cloud provider not only provides the hardware, but they also provide a toolkit and a number of supported programming languages to build higher level services (i.e. software applications that are made available as part of a specific platform). The users of PaaS are typically software developers who host their applications on the platform and provide these applications to the end-users.

Examples of PaaS providers include Google App Engine [16], Microsoft Windows Azure [17] and Force.com [18].

### C. Software as a Service (SaaS)

Third and last layer is the SaaS. The SaaS customer is an end-user of complete applications running on a cloud infrastructure and offered on a platform on-demand. The applications are typically accessible through a thin client interface, such as a web browser. The customer does not control either the underlying infrastructure or platform, other than application parameters for specific user settings.

Examples of SaaS providers include Salesforce.com [18], Rackspace [19] and SAP Business ByDesign [20].

### III. TYPES OF CLOUD

The different infrastructure deployment models are distinguishing by their architecture, the location of the datacenter where the cloud is realized, and the needs of the cloud provider's customers. Cloud can be of many types [1] [2] [3] [7]. Following are the categories of cloud:

- Public Cloud
- Private Cloud
- Hybrid Cloud
- Community Cloud

### A. Public Cloud

Public clouds are provided by third party i.e. all the facilities are provided by some third party with all information technology resources residing in their datacenter, such cloud is shared among many customers and all customers have equal chance of utilizing those resources.

### B. Private Cloud

Private clouds are provided by an organization to its employees or subordinate organizations. All the IT resources are residing in the organization datacenter who owns the cloud and the resources are shared only among the employees of the organization or the organizations working under the main company who owns the clouds.

### C. Hybrid Cloud

Hybrid cloud is the combination of public and private clouds. Hybrid cloud is basically used when the service providers or the cloud owners do not want to invest in datacenters and to extend cloud services to accommodate more and more user's requests. Thus public cloud providers borrow facilities of private cloud to extend services.

### D. Community Cloud

Community clouds are also provided by third party and they service organizations having similar interest e.g. hospitals. Specific software are developed which could be used by all hospitals under community cloud umbrella.

### IV. BENEFITS OF THE CLOUD

The Clouding computing has many more advantages, but from the following section one can understand how important cloud computing is, and how it will affect the computing [7] [8] [10] [12].

### A. Reduce Hardware Cost.

All the work is done in the cloud and it will reduce the cost for purchasing high cost equipment for the organization having thousands of employees. The employees only need a terminal to connect to the cloud in order to perform most of the computation. Not only the personal computer for the employees but it also reduces the cost which is incurred on the purchase of high cost server machines.

B. *Reduce Software Costs.*

The proprietary software is no longer needed to purchase. The amount is paid to the cloud provider as when it is needed to use the software instead of buying high cost software. It also reduces the software cost which needed to run and managed any organization's server.

C. *Maintenance and Upgrading Cost.*

It is possible for the employers to quickly remove associated computer costs when the number of employees is reduced. It is easy to migrate, or upgrade the current operating system, hardware etc. with a new one, because the organization only needs to pay for the services which they want to upgrade instead of investing again and purchasing the high cost software and hardware.

D. *Preconfigured System.*

It is easy to add new computers right away, and avoid the higher cost of adding a computer through the conventional route which could take hours to setup, install, and configure the applications your company uses.

E. *Data Locality.*

One the benefit of the cloud computing is that although the organization is not aware of the physical location of the data, but they view the data to present in one location. So if an organization has multiple offices then the data can be seen to be present on one single location.

F. *Back – Up Facility.*

Cloud computing provides an automatic data backed – up facility as opposed to a desktop computer or notebook computer does set to automatically save important data on server.

G. *Reduces the Risks of Theft.*

As the data resides on the cloud, so, if a company's notebook or any other computing equipment is stolen, then there will be less chances of losing the company's proprietary and sensitive data, and it will also reduces the chances of greater financial impact.

H. *Availability, Sharing and Collaboration.*

If a company has all its important data or computation on the cloud, then it is very easy to access your data from anywhere if you have only a computer terminal and it is connected with the internet. Similarly it also allows the participant to share and work on the same instance of the data. It is easy for the organization to expand its branches.

I. *Cloud Benefits for Short Term Projects.*

If a company have a short term project, and it needs to work on specific software such as Microsoft Project, then cloud computing is the best available solution for them. It is easy to use the software for the duration of the project instead of buying the software.

J. *Free up Office Space.*

The elimination of the onsite large servers and related networking equipment from the organization due to the adoption of the cloud computing, the space is freed up.

K. *Reduce Cost of Electricity.*

Due the reduction in the size of servers and other related hardware equipment, this is present at the organization's location, so the consumption of the electricity is reduced. Therefore it decreases the amount of the bill.

L. *Self Service Concept.*

The Cloud computing completely is based on self service concept. Customer or service provider is responsible for the services he uses or provides, no administrator is available to configure the resources or provision/deprovision the resources.

M. *Faster Time to Market.*

Since no investment is done in hardware so provider only need to concentrate on services he has to offer to the community, in this way software is brought to market at much faster rate.

N. *Quality of Service.*

User gets services from cloud using internet so he/she is not responsible for the underlying infrastructure of the service i.e. if the system is properly working or any system is failed completely or faulty, every such concern is managed by the could itself by its distributed nature so user get a quality service.

O. *Business Flexibility.*

Since many businesses depend on internet to advertise its products and online sale thus cloud computing provides flexibility to such businesses to grow up in peek seasons and shrink when demand is low.

V. CLOUD COMPUTING RESEARCH CHALLANGES

Besides having all these advantages, cloud computing is yet not safe from threats. There are some issues that are creating bottleneck in making cloud computing concept widely accepted [9] [10] [11] [12]. Besides having all these advantages, cloud computing is yet not safe from threats. There are some issues are creating bottleneck in making cloud computing concept widely accepted. The challenges are follows:

- Portability and Interoperability (Lock – In)
- Development of New Architecture
- Availability of Service and Limited Scalability
- Lack of Standards

- Security and Privacy
- Reliability
- Governance and Management
- Metering and Monitoring
- Energy Management in Cloud
- Denial of Service (DoS)

A. *Interoperability and Portability (Lock – in)*

Interoperability and portability present another open research problem for the researcher. Interoperability is the way how different clouds would communicate. It refers to the ability of customers to use the same parameters-management tools, server images etc- with a variety of cloud computing providers and platforms e.g. Amazon and Google are two clouds. Using the same image of Windows from Amazon on Google without any change is called interoperability. This would require Google to understand Amazon language.

Portability refers to the ability to move application and its data from one cloud to another. Portability could be achieved by removing dependencies on the underlying atmosphere. A portable components (application, data) could be moved and reused regardless of the provider, platform, operating system, location, storage etc without being modified e.g. if the old cloud environment is Windows and new cloud environment is Linux then an application running on old cloud would be able to run on new cloud without being changed is called portability.

B. *Development of New Architecture*

Presently, almost all of the cloud computing services are implemented in large commercial data centers and they are operated in old centralized manner. This design has its benefits i.e. Economy of scale and high manageability, yet it has some limitations i.e. High energy consumption and initial cost of investment. Most of the researchers have an inclination towards using voluntary resources to host cloud applications. This model of cloud computing in which using voluntary resources, or a mixture of both dedicated and voluntary resources are very economical and it suits such applications as scientific computing. However, despite its advantages, yet this architecture has open research challenges as well, which are heterogeneous resources management, incentive scheme for such architecture.

C. *Availability of Service and Limited Scalability*

Since many systems have been crashed on cloud like Amazon so using only one CCSP services can result in a drawback as when a shutdown event happens on a cloud the service disappears and user cannot find that service.

CCSP promise to provide infinite scalability for customer but due to the fact that millions of users are now migrating to cloud computing so such promise is not fulfilled.

The challenge of availability and scalability presents another research area for the researcher to find an optimum solution for these problems.

D. *Lack of Standards*

Every cloud provider has his own standards and user is not given any comparative performance measurement facility by which he can compare standards and performance of different clouds using some cost per service metric. It is still needed that cloud computing should be standardized, and a lot of research work is still needed to meet the required level of standardization.

E. *Security and Privacy*

The main hurdle in the fast adoption of cloud is the security concerns of the customers. Although due to the presence of modern techniques of security the chances of security flaws are reduced but still, when worms and hackers attack a system, mayhem is created within a few hours. It is necessary that the applications and architectures should be secluded and the mechanism of security must be apposite, surfacing and adoptive. A lot of research work is done, going on and still needed to be carried out in the area of cloud security. Trust and Privacy are some other potential areas of research in cloud computing.

F. *Reliability*

Availability of connection to cloud network is again an issue. User is not sure if he will remain connected to cloud network and keep on doing his work at any time as connections do break. The connections to cloud services are secure or not and the migration of data to cloud computing is in safe environment and as per needed speed or not. Cloud itself is reliable enough to be migrated to? So reliability is another challenge yet to be resolved.

G. *Governance and Management*

Many organizations started providing cloud services using their own data centers, thus trying to govern and bring monopoly in cloud computing. Governments, organizations and users must need to work together to resolve this issue.

H. *Metering and Monitoring*

Organizations using cloud services must monitor the performance of services. Services providers must provide means to measure and monitor their services across standard parameters.

I. *Energy Management in Cloud*

The primary requirement of the cloud computing is the management of heterogeneous resources across a distributed computing environment. It is obvious that from the user point of view all these resources are "on all the times". If this is the case, then, it is highly inefficient in terms of the requirement for the energy consumption. A lot of research has been carried out in developing energy efficient equipment and utilize these equipments in building data centers to be energy efficient. On the other hand, fewer efforts have been put to model and exhibit a potential which allows various, distributed clouds infrastructure to use a policy which demonstrate to be as energy efficient as possible. The research work is needed to be carried out in the area of virtualization not only in system but network resources as well to minimize the energy consumption.

*J. Denial of Service*

Another burning issue and a challenge that is faced by the researcher working in the area of cloud is the denial of service (DoS) in cloud computing. As a matter of fact cloud offers the allocation of resources dynamically, so what will be the response of the cloud when it is under a heavy denial of service attack? Is it necessary to built a DoS protection into cloud, or it will be handled on the internet level as it is dealt presently. This also poses another challenge for the researcher.

## VI. CONCLUSION

Cloud computing is yet in its initial stages but the advantages it has brought along are tremendous. Due to the lack of standards, users and computing industry is still reluctant to fully accept cloud computing concept. Burden lie on researcher's shoulders to bring about new architectures which would help in removing issues and threats related to cloud computing so that majorities can take full advantage of its wonderful features.